\documentclass[]{spie}  

\usepackage{amsmath,amsfonts,amssymb}
\usepackage{graphicx}
\usepackage[colorlinks=true, allcolors=blue]{hyperref}
\usepackage{wrapfig}
\usepackage{lscape}
\usepackage{rotating}
\usepackage{epstopdf}

\title{Analysis of active optics correction for a large honeycomb mirror}

\author[a,b]{Solvay Blomquist}
\author[a]{Hubert Martin}
\author[a,b]{Hyukmo Kang}
\author[a]{Rebecca Whitsitt}
\author[b]{Kevin Derby}
\author[b]{Heejoo Choi}
\author[a]{Ewan S. Douglas}
\author[a,b]{Daewook Kim}
\affil[a]{Steward Observatory - The University of Arizona \linebreak 933 N Cherry Ave, Tucson, AZ 85721 \linebreak}
\affil[b]{Wyant College of Optical Sciences - The University of Arizona \linebreak 1630 E University Blvd, Tucson, AZ 85721}

\authorinfo{Further author information: (Send correspondence to H.M.)\\H.M. E-mail: hmartin@arizona.edu}

\begin{document} 
\maketitle

\begin{abstract}
In the development of space-based large telescope systems, having the capability to perform active optics correction allows correcting wavefront aberrations caused by thermal perturbations so as to achieve diffraction-limited performance with relaxed stability requirements. We present a method of active optics correction used for current ground-based telescopes and simulate its effectiveness for a large honeycomb primary mirror in space. We use a finite-element model of the telescope to predict misalignments of the optics and primary mirror surface errors due to thermal gradients. These predicted surface error data are plugged into a Zemax ray trace analysis to produce wavefront error maps at the image plane. For our analysis, we assume that tilt, focus and coma in the wavefront error are corrected by adjusting the pointing of the telescope and moving the secondary mirror. Remaining mid- to high-order errors are corrected through physically bending the primary mirror with actuators. The influences of individual actuators are combined to form bending modes that increase in stiffness from low-order to high-order correction. The number of modes used is a variable that determines the accuracy of correction and magnitude of forces. We explore the degree of correction that can be made within limits on actuator force capacity and stress in the mirror.  While remaining within these physical limits, we are able to demonstrate sub-25 nm RMS surface error over 30 hours of simulated data. The results from this simulation will be part of an end-to-end simulation of telescope optical performance that includes dynamic perturbations, wavefront sensing, and active control of alignment and mirror shape with realistic actuator performance.
\end{abstract}


\section{INTRODUCTION}
\label{sec:intro}  

The scope of this study is to demonstrate control of the shape of a large honeycomb mirror as it is perturbed by changing thermal gradients in a dynamic environment. The method of active optics correction has been developed and exercised for other honeycomb mirrors fabricated at the Richard F. Caris Mirror Lab (RFCML) for ground-based telescopes such as the Large Binocular Telescope (LBT) and the MMT. The mirror shape is controlled by actuators that apply force to the rear surface of the mirror. A modal correction is applied, using a limited number of bending modes, or combinations of the individual actuator forces, to alter the shape of the mirror surface. \cite{10.1117/12.176238, 10.1117/12.550464, 10.1117/12.319262,10.1117/12.672454} The bending modes are ordered from low to high spatial frequency correction. Using more bending modes reduces the residual surface error but increases the range of actuator forces. Using more modes also increases the uncertainty in actual performance relative to simulated performance. We generally want to use enough bending modes to meet the requirements for wavefront error, and not significantly more. The simulations are valuable in determining a reasonable number of modes for the control system.  

The input wavefront errors for this simulation are representative of the errors that might be due to thermal evolution of a spacecraft on orbit. These are a single statistical realization of a hypothetical observatory\cite{ewan_talk}, and serve as a proof of concept. Future work will more fully explore the sensitivity of the bending modes to different wavefront time series (e.g. as done in Lyon and Clampin\cite{lyon_space_2012}). In this study, we consider the time series of input wavefront error and assume ideal, near real-time measurement of the wavefront error. Residual error therefore represents the limitations of the active control of mirror shape with a given number of bending modes and is not limited by sensor noise or phase retrieval algorithms used\cite{kevin_talk}.

\section{Active Support and Correction System}
\label{sec:sections}

Active optics generally includes control of alignment of the secondary mirror (M2) and control of the shape of the primary mirror (M1), based on feedback from a wavefront sensor. In this paper, we focus on controlling the shape of the primary mirror. This study is part of a larger simulation of the full active optics system for a large telescope, including wavefront sensing, alignment of M2 and bending of M1\cite{kevin_talk}. For the analysis presented in this proceeding, we assume tilt has been corrected by pointing the telescope, and focus and coma have been corrected by aligning M2. All remaining wavefront error will be addressed by bending M1 with its actuators. 

We consider a case where the dominant wavefront errors come from temperature gradients. Gradients in the telescope structure cause pointing error and misalignment of M2. Gradients in M1 cause thermal distortion of M1.  

The M1 support system includes six hard points that define its position in six degrees of freedom, and 166 axial force actuators that control its shape. Forces on the hard points are sensed and transferred to the force actuators, so the hard points apply minimal force at all times\cite{10.1117/12.550464}. The actuator force capacity exceeds $\pm$ 1000 N.

\section{Bending Mode Calculation and Fit}

The simulated correction of M1 surface error involves fitting a set of bending modes to the surface error. The bending modes, which are linear combinations of actuator influence functions, form a basis set for the full range of shape change that can be made with the 166 actuators. They are ordered from low to high spatial frequency, or equivalently from most flexible to stiffest. Figure 1 shows several bending modes and illustrates the progression in spatial frequency. Using bending modes allows a modal solution, where the number of modes can be selected to optimize performance.  

\begin{figure} [h!]
\begin{center}
\begin{tabular}{cc}
    \includegraphics[height=6cm]{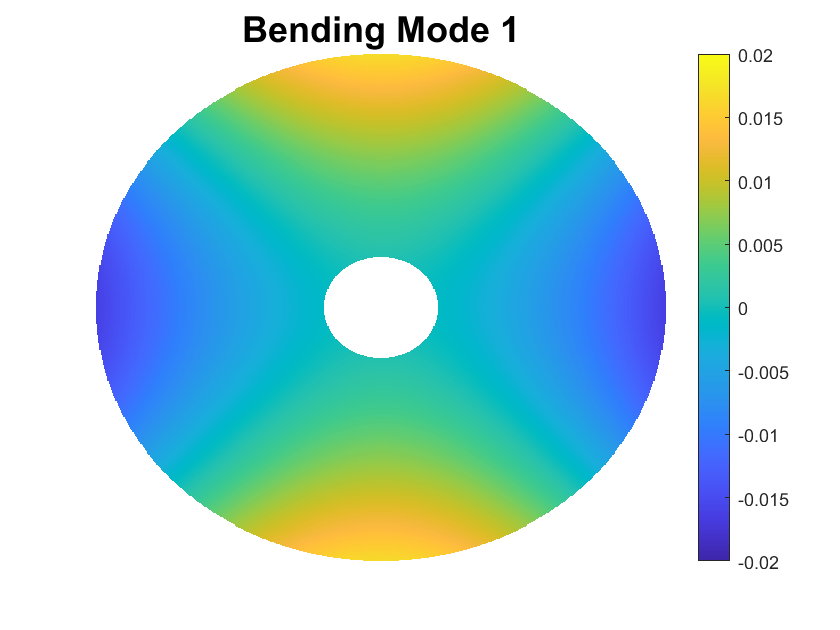} & \includegraphics[height=6cm]{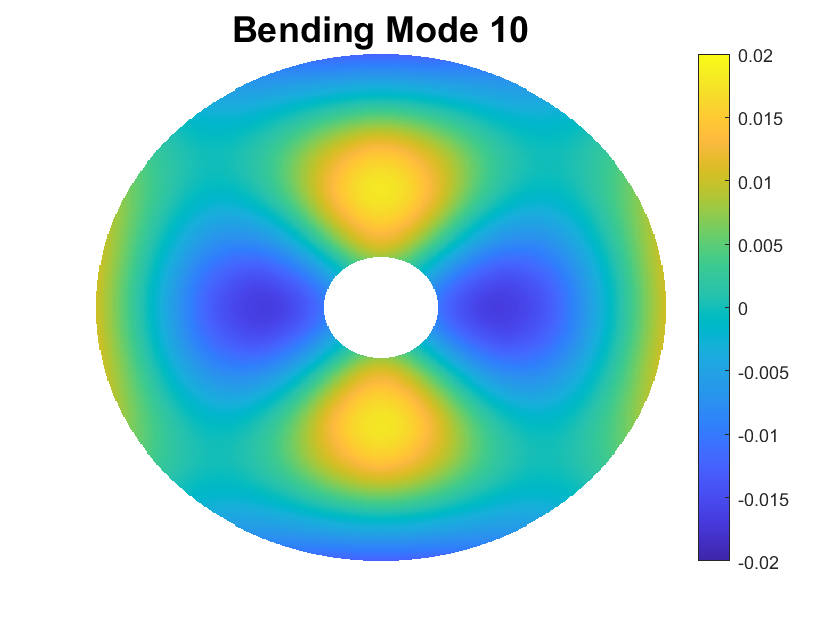} \\ 
    \includegraphics[height=6cm]{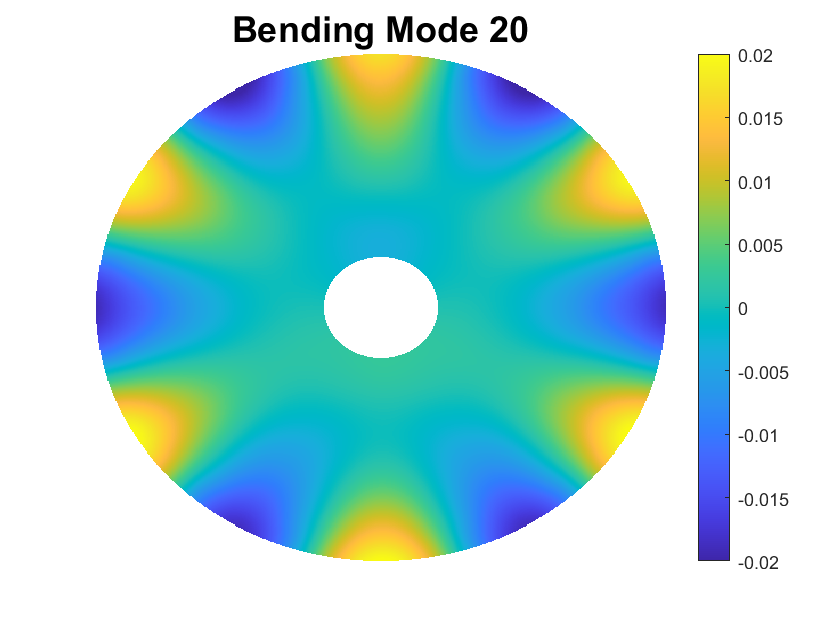} & \includegraphics[height=6cm]{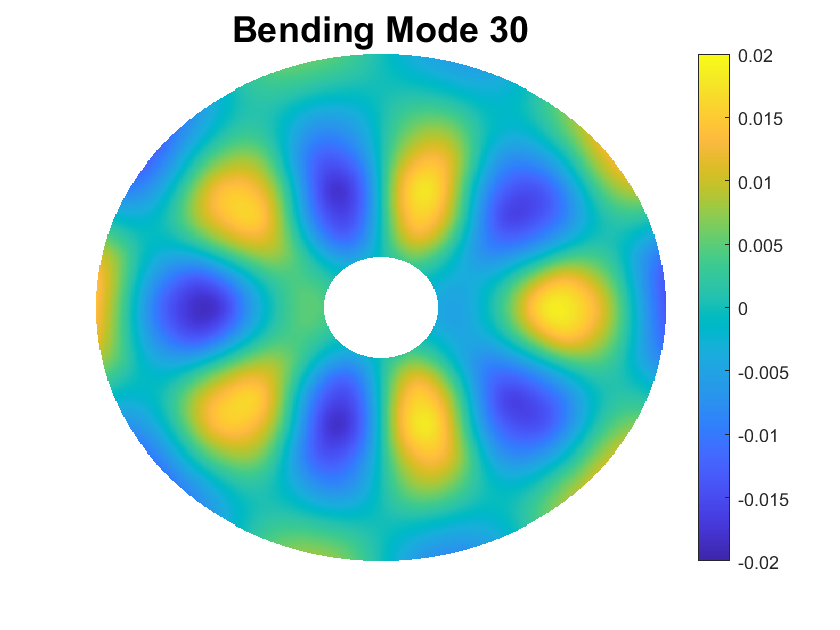} \\
    \includegraphics[height=6cm]{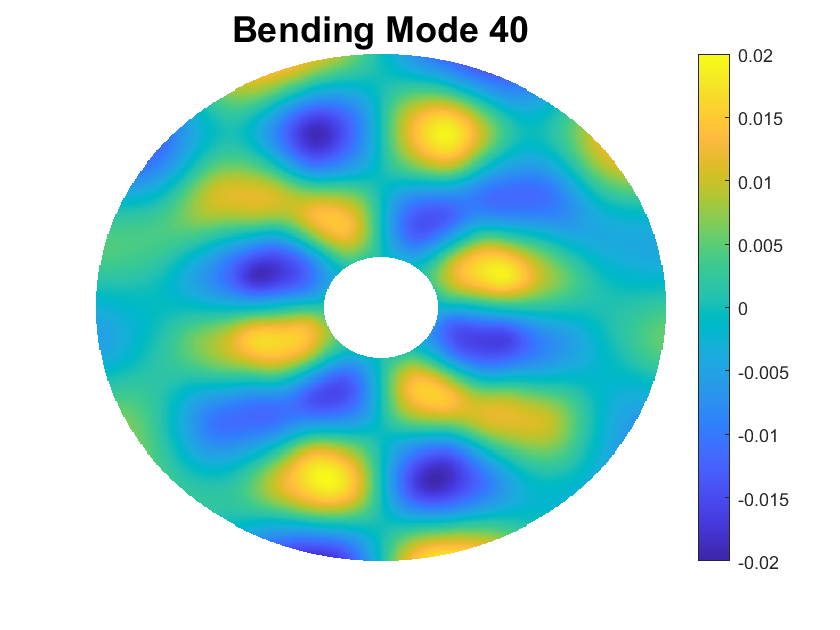} & \includegraphics[height=6cm]{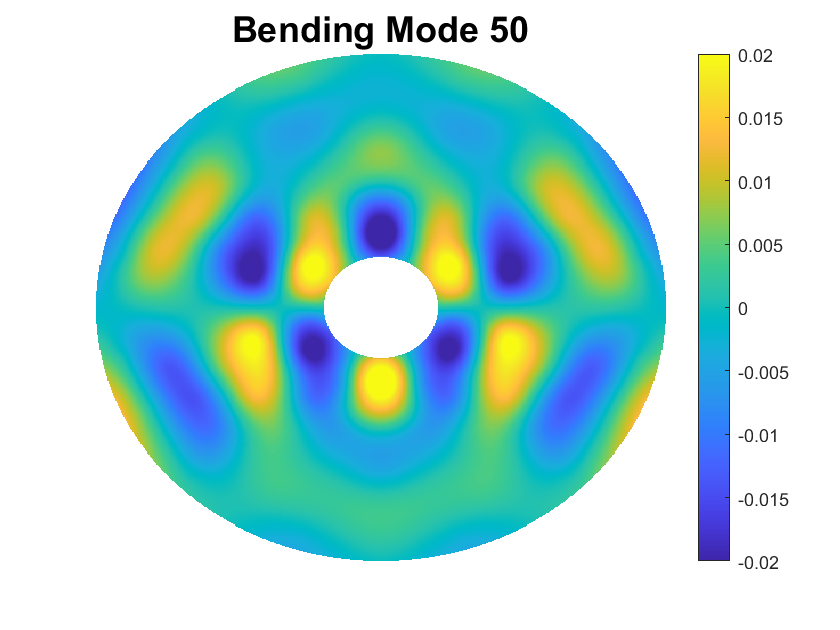} \\ 
\end{tabular}
\end{center}
\caption[example] 
{ \label{fig:video-example} 
Selected bending modes of the honeycomb primary mirror with 166 force actuators, illustrating the progression from low to high spatial frequency and stiffness. The modes are normalized to identical RMS surface deflection. }
\end{figure} 

We start by using the finite-element model to compute the  actuator influence function, i.e., the shape change for a unit change in actuator force, for each of the 166 actuators. Laboratory measurements of other large honeycomb mirrors confirm that the measured influence functions agree well with those computed with the finite-element model. \cite{10.1117/12.550464, 10.1117/12.319262} We then use singular value decomposition to compute the bending modes, the corresponding sets of actuator forces, and the mode flexibilities. \cite{Press_1988} The mode flexibility is the ratio of RMS surface deflection to RMS force over the 166 actuators.

We simulate the correction of surface errors by fitting a given number of bending modes to the initial surface error. The fit represents the part of the initial error that can be corrected with the given modes. The residual error represents the part that cannot be corrected. 

Figure 2 shows the flow of data through the analysis pipeline for the current study. It starts with a map of M1 surface error, which may come from a thermal model or, for the data presented here, from simulated data that are representative of thermal distortion of M1. \cite{ewan_talk} For consistency with the simulation of the full active optics system, we use a 231-term Zernike polynomial fit to represent the M1 surface error. We use Zemax to compute the resulting wavefront error in the focal plane. (In the full simulation, we also include here the wavefront error due to misalignment of all the telescope optics.) The active optics correction is based on the measured wavefront error in the focal plane. A second 231-term Zernike fit to the corrected surface is fed back to the full simulation as M1's new surface shape. \cite{kevin_talk} Wherever a Zernike fit is used, we keep track of the fitting error (data minus fit) and consider it along with other analysis errors when we estimate the full system wavefront error.  \cite{heejoo_talk}

\clearpage

\begin{figure}[h!]
    \begin{center}
    \includegraphics[height=5cm, width=11cm]{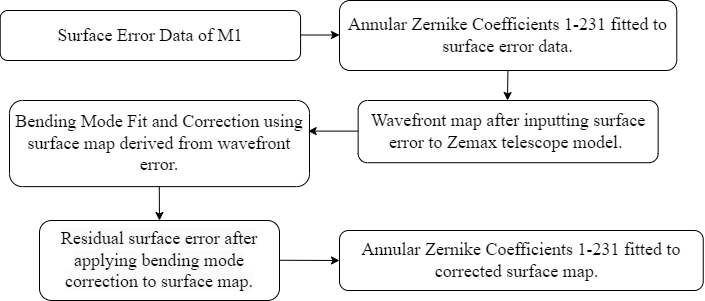} 
    \end{center}
    \caption{Flow chart for simulating the active optics correction of M1. M1 surface errors are represented by Zernike fits, to maintain compatibility with the simulation of the full active optics system. }
\end{figure} 

\section{Numerical Modeling and Simulation Results}

\begin{figure}[h!]
    \begin{center}
    \includegraphics[height=8cm]{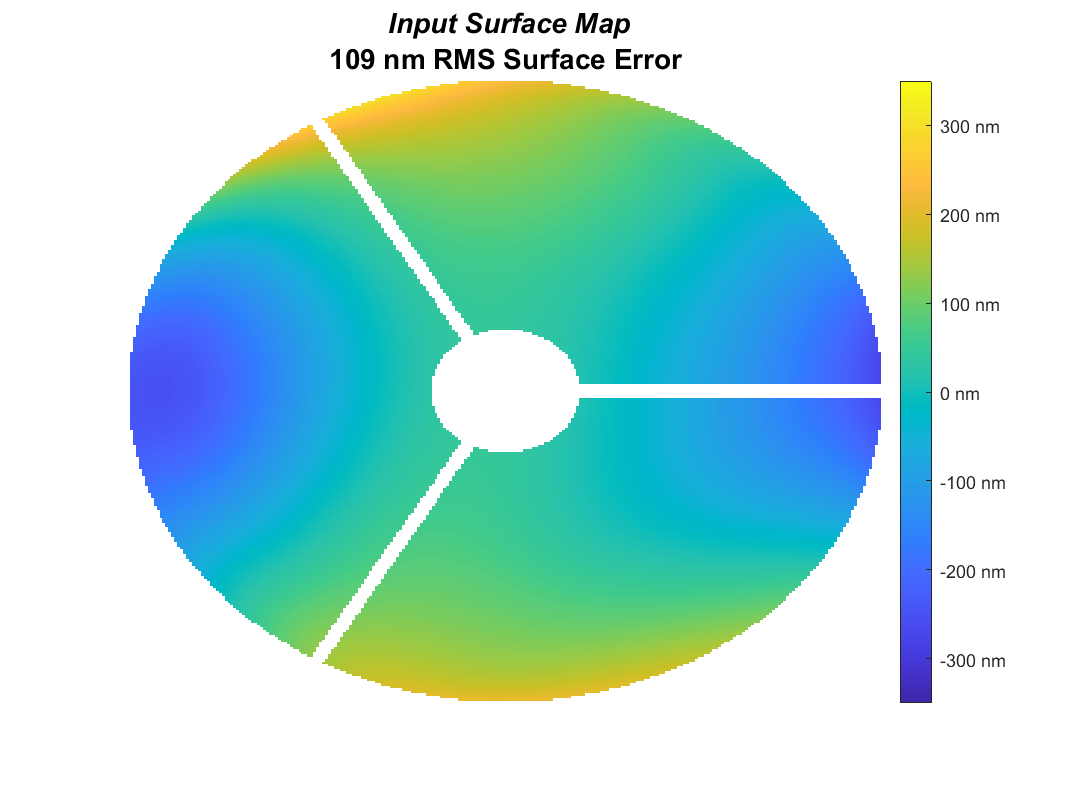} 
    \end{center}
    \caption{Surface error in nm taken from time series of simulated data. This is the input to the active optics correction. This map represents the point in the time series with the highest RMS surface error, 109 nm. The 3-legged obscuration is a generic representation of a tripod to support M2.}
\end{figure}

As an example of the simulated correction of M1, we start with the uncorrected surface error shown in Figure 3. The dominant error is astigmatic and will be corrected with a single bending mode, but significant errors will remain. The number of modes used in the correction is an important variable. Figure 4 illustrates how effective the correction is for 1, 14 and 68 modes. In this case, while a single mode removes most of the RMS surface error, at least 14 modes are needed to correct the large-scale structure and approach 10 nm RMS surface error. We include the 68-mode correction in order to explore the potential capability to correct thermal distortion. Even with 68 modes, extreme actuator forces are well within the $\pm$ 1000 N capacity, and the stress in the glass remains at a safe level, but it's not yet clear that we can control as many as 68 modes with good stability and accuracy. 

\clearpage

\begin{figure} [h!]
\begin{center} 
\begin{tabular}{cc}
   \includegraphics[height=6cm]{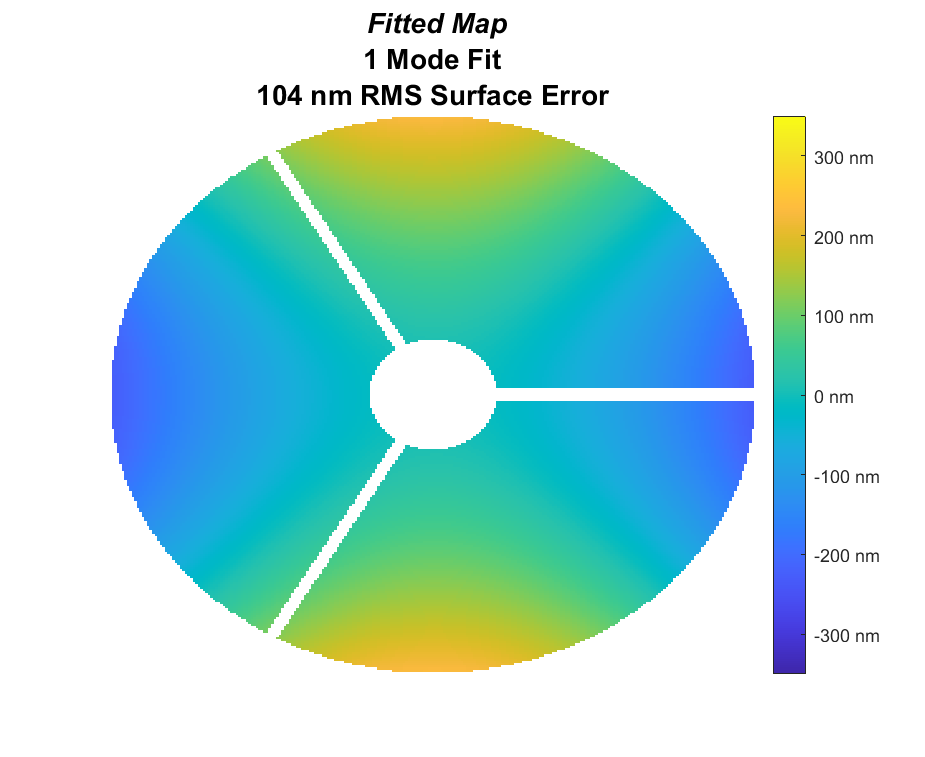}  & \includegraphics[height=6cm]{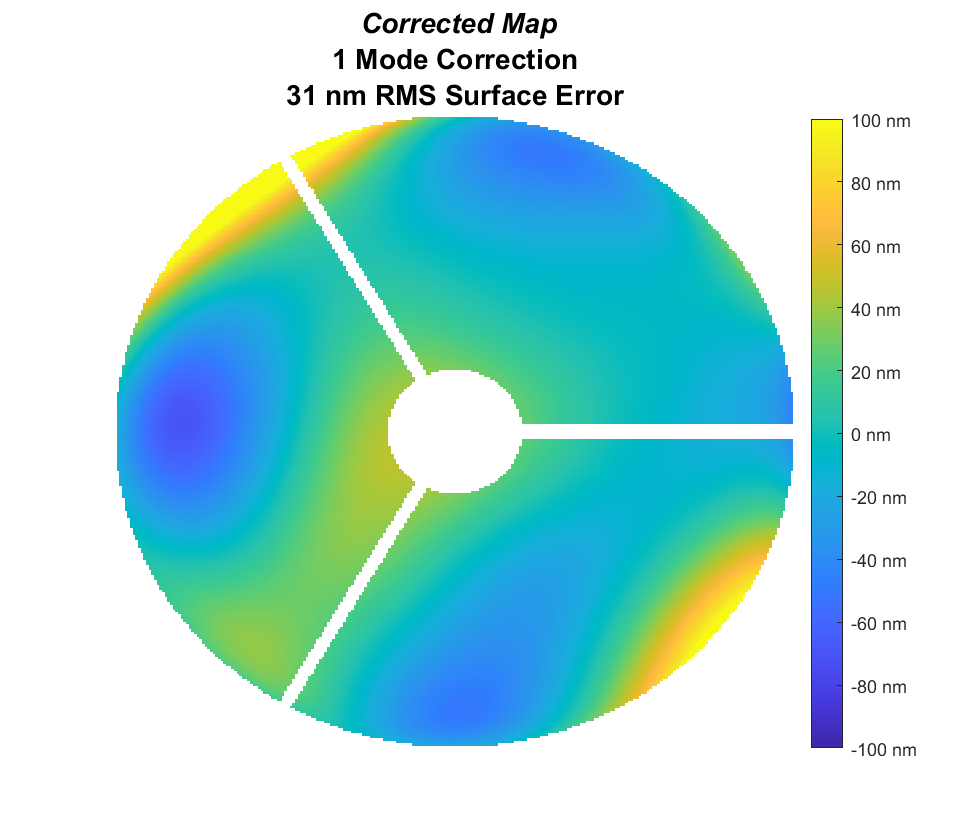} \\
   \includegraphics[height=6cm]{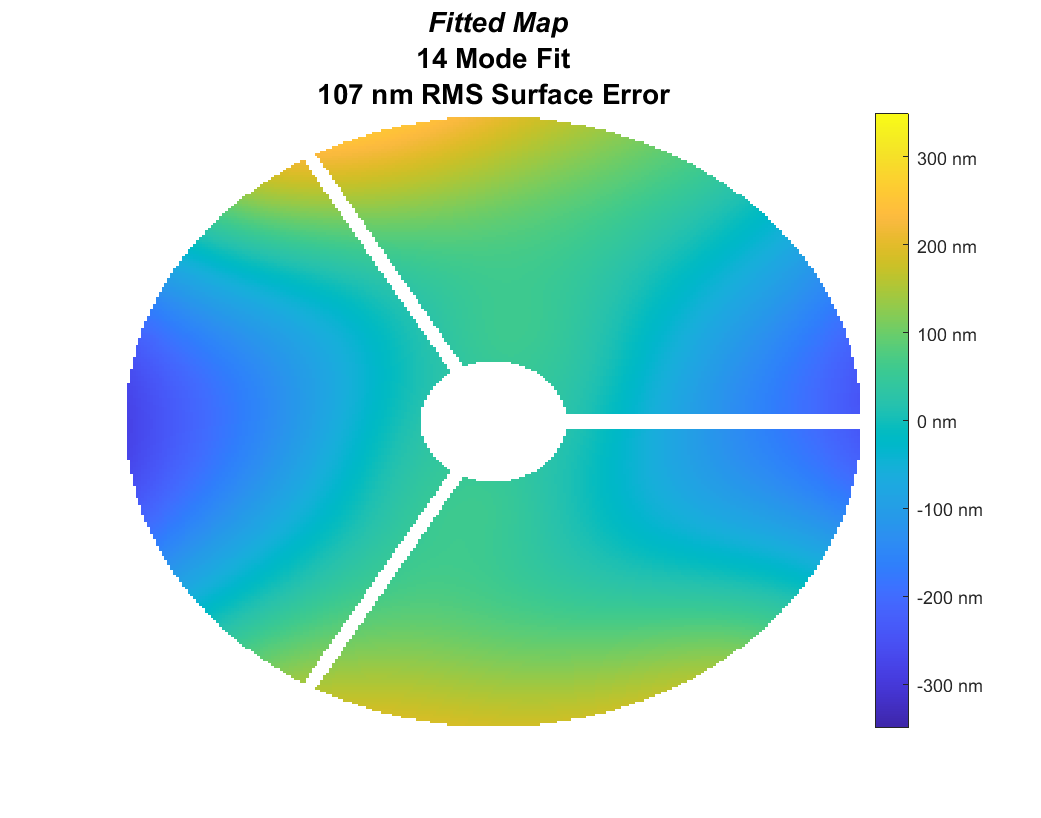} & \includegraphics[height=6cm]{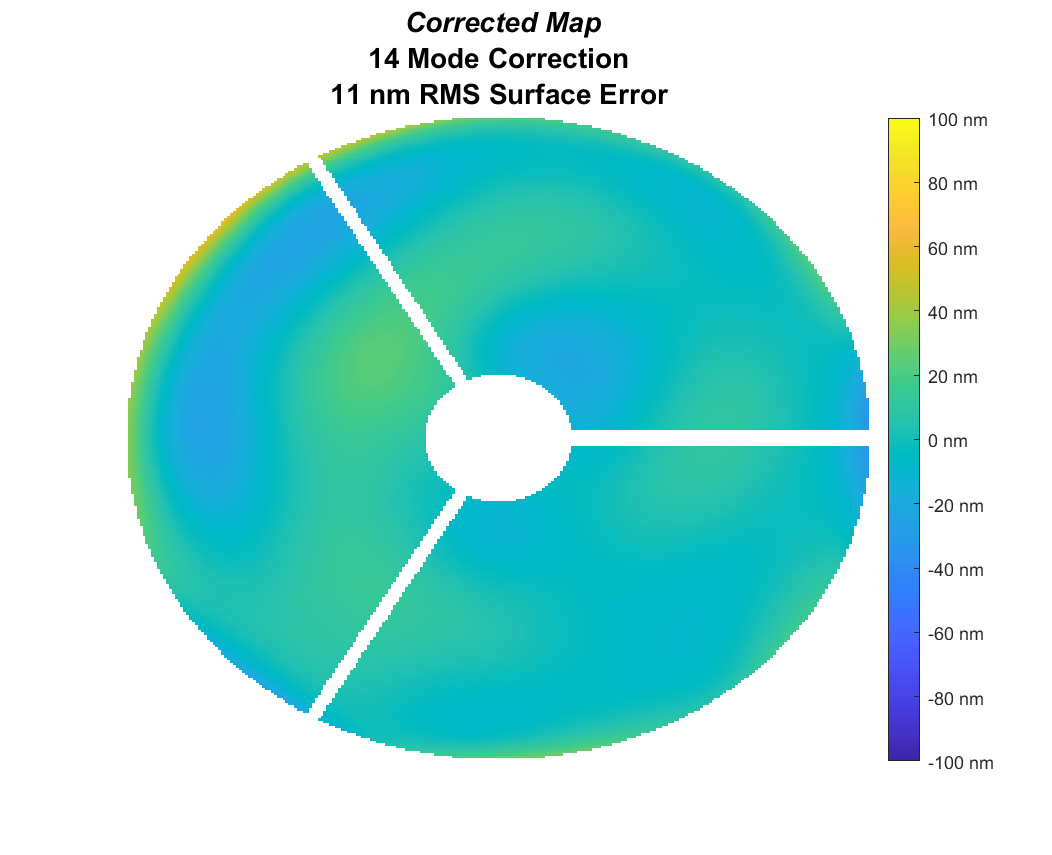} \\
   \includegraphics[height=6cm]{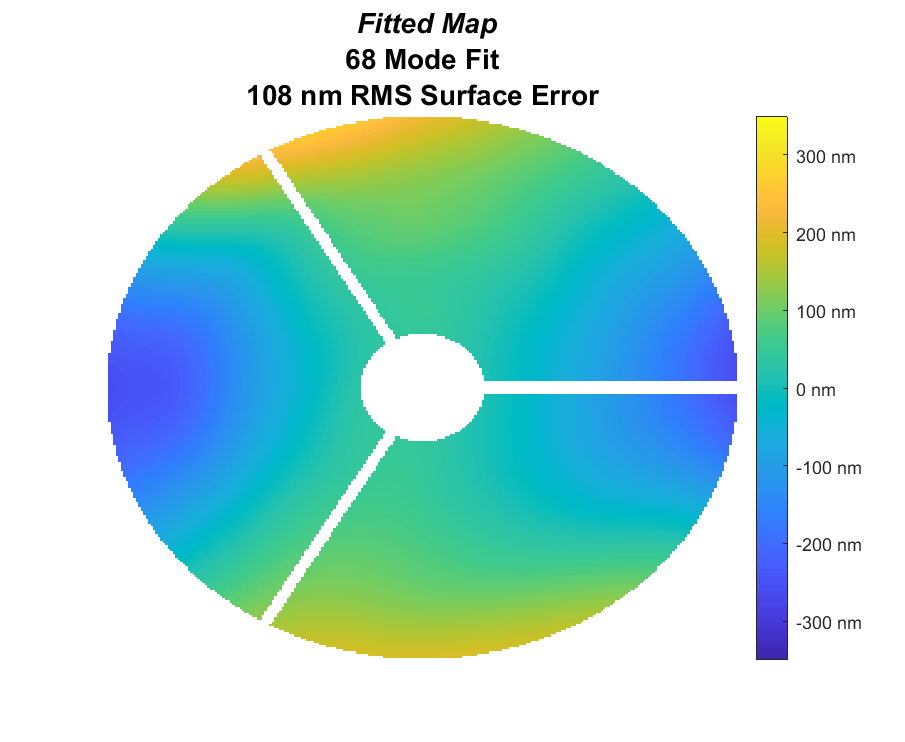} & \includegraphics[height=6cm]{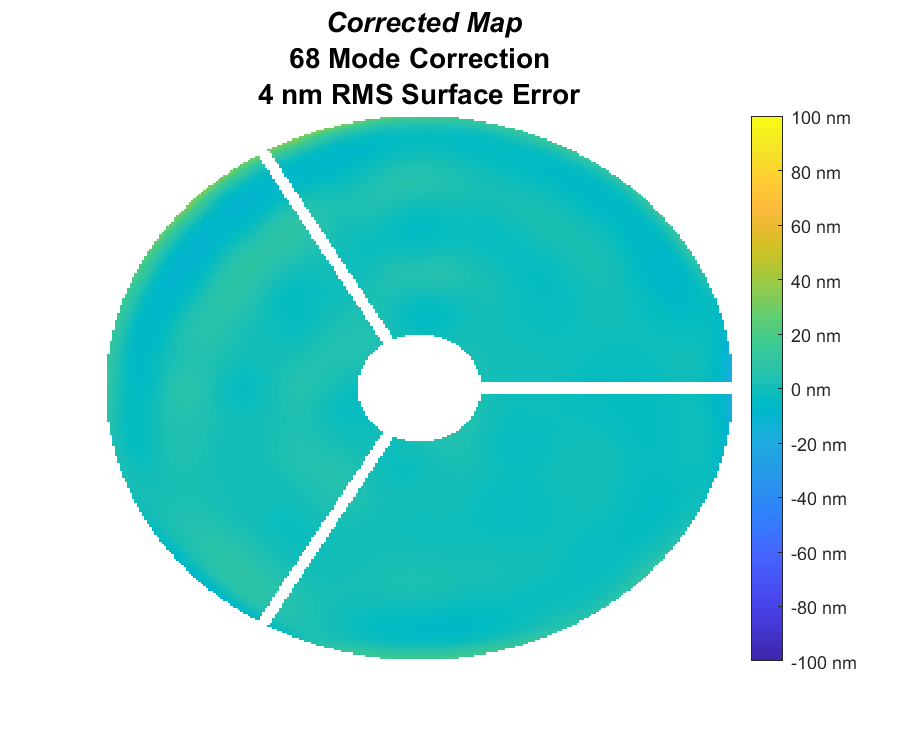}
\end{tabular}
\end{center}
\caption[example] 
{ \label{fig:video-example} 
Given an input surface error as shown in Figure 3, different numbers of bending modes are fit and subtracted from the initial map to obtain a corrected map. Color bars are labeled in nm of surface error.}
\end{figure} 

\clearpage

For the simulated data, even the 14-mode correction reduces thermal distortion to the point that it will not be the dominant wavefront error. Manufacturing error due to limitations in polishing and in-process measurements will leave mid-scale structure at the level of 10-15 nm RMS surface error.

Figure 5 shows a 30-hour time series of simulated data, including RMS surface error for the input (uncorrected) surface map and the corrected surface using 14 and 68 bending modes. The input map has had tilt, focus and coma removed, corresponding to pointing the telescope and aligning M2. The 14-mode correction meets the goal (12.5 nm RMS surface error \cite{heejoo_talk}) for most of the 30 hours. The 68-mode correction comfortably meets the goal for the full period, and remains consistent with the actuator capacity and the glass stress limits. Further analysis, including simulation of the full active optics system, will guide the choice of number of modes to correct. This parameter can be changed dynamically as conditions change.

\begin{figure}[h!]
\includegraphics[width=7in]{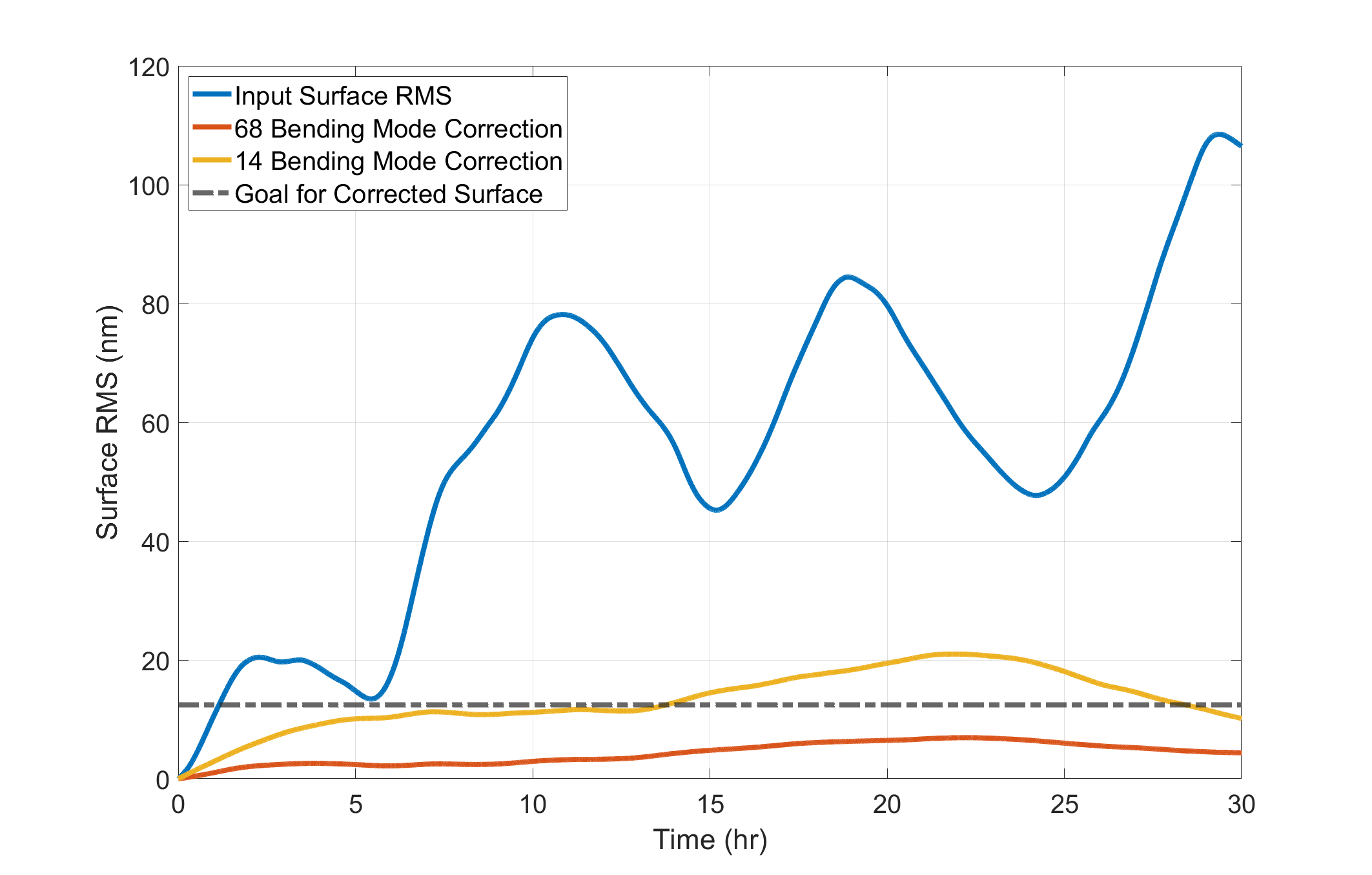}
\begin{center}
\caption[example] 
{ \label{fig:video-example} 
RMS surface error for the uncorrected surface and for the surface after correction with 14 and 68 bending modes, for a 30 hour time series of simulated data. The goal for the corrected surface RMS is at 12.5 nm.}
\end{center}
\end{figure} 

Figure 6 shows how the corrected surface improves and the actuator forces increase as we increase the number of bending modes in the correction. The input map for this plot is the map shown in Figure 3. The RMS surface error improves dramatically after correction with the first mode, which happens to match the dominant astigmatism in the input surface error. It continues to improve rapidly as modes 2-14 are added to the correction. The RMS actuator force remains low, less than 12 N, through 14 modes. Including more modes, up to 68, slowly reduces the RMS surface error by another factor of 2.5 while the RMS actuator force increases steadily by a factor of 8. While the actuator forces and glass stress remain acceptable, we expect that a real 68-mode correction will not behave nearly as well as the simulated correction. At best, multiple iterations would be needed to achieve the simulated result.

\clearpage

\begin{figure}[h!]
\begin{center}
\includegraphics[height= 10cm]{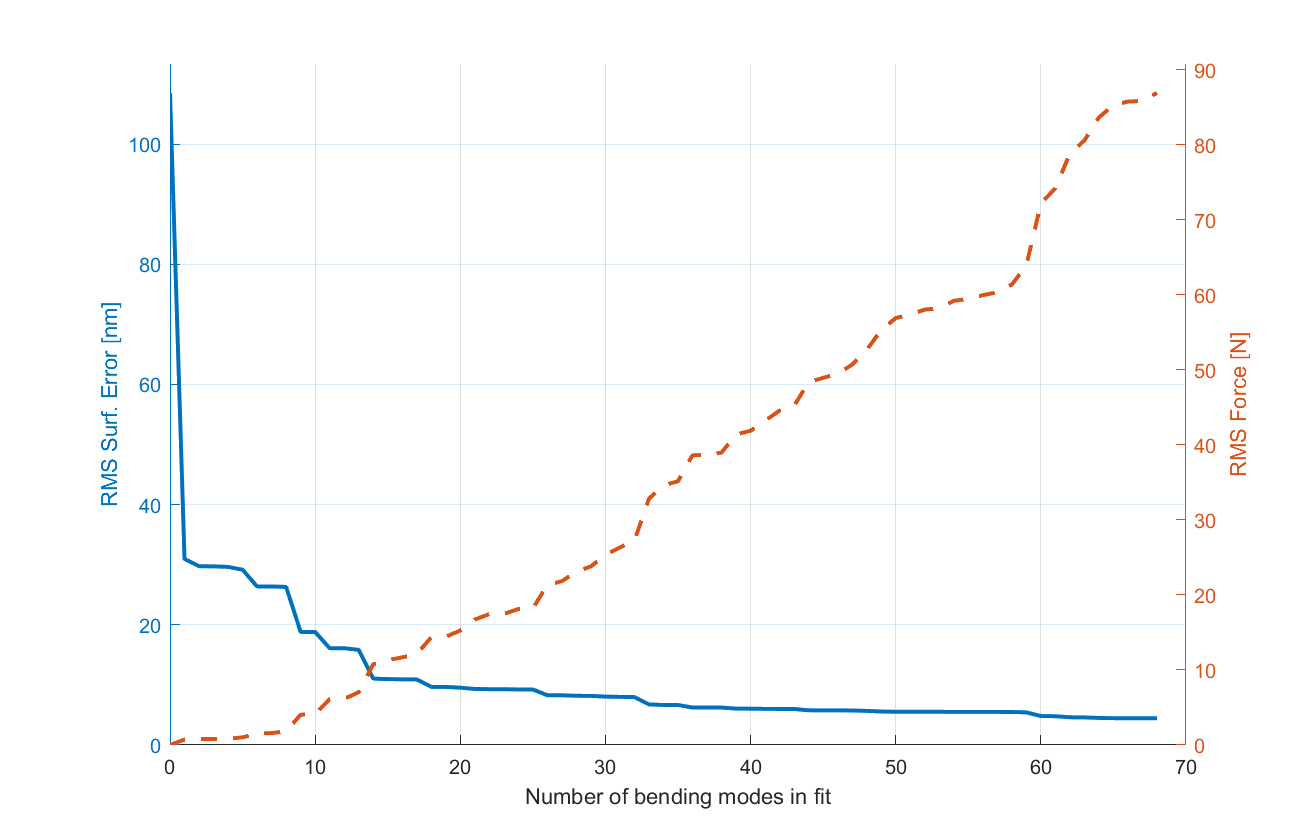}
\end{center}
\caption[example] 
{ \label{fig:video-example} 
RMS surface error and RMS force over the 166 actuators, as a function of the number of bending modes used in the correction. The input surface for this graph is the map shown in Figure 3.}
\end{figure}

\section{Conclusions}

This study demonstrates a viable method of correcting for thermal deformation of a large honeycomb primary mirror in space using the mirror modes used for ground-based control of RFCML mirrors. Results to date indicate that the mirror support system with 166 force actuators has excellent authority to bend out thermal deformations with acceptable actuator forces and glass stress. The examples we show illustrate how the accuracy of the corrected surface and the magnitude of actuator forces depend on the number of bending modes used in the correction. We can use the simulation to determine an optimum number of modes for different conditions and generate an envelope of the allowable wavefront error perturbations. We continue to refine the full active optics simulation to include realistic wavefront measurements, actuator performance, and other sources of wavefront error.

\section*{ACKNOWLEDGEMENTS}

Portions of this research were supported by funding from the Technology Research Initiative Fund (TRIF) of the Arizona Board of Regents and by generous philanthropic donations to the Steward Observatory of the College of Science at the University of Arizona.

\bibliography{report} 
\bibliographystyle{spiebib} 

\end{document}